\documentclass[superscriptaddress,reprint]{revtex4-2}
\usepackage{amsmath}
\usepackage{graphicx}
\usepackage{color,soul}
\usepackage[normalem]{ulem}
\usepackage{cancel}
\usepackage{dcolumn}
\usepackage{bm}
\usepackage{hyperref}
\usepackage[mathlines]{lineno}

\begin{document}

\preprint{APS/123-QED}

\title{Stimulated Brillouin scattering at 1 nm$^{-1}$ wavevector\\ by extreme ultraviolet transient gratings}

\author{Danny Fainozzi}
 \email{danny.fainozzi@elettra.eu}
\author{Laura Foglia}
\affiliation{Elettra-Sincrotrone Trieste, SS 14 km 163,5 in AREA Science Park, 34149, Trieste, Italy.}
\author{Nupur N. Khatu}
\affiliation{Elettra-Sincrotrone Trieste, SS 14 km 163,5 in AREA Science Park, 34149, Trieste, Italy.}
\affiliation{Department of Molecular Sciences and Nanosystems, Ca’ Foscari University of Venice, Venice, Italy.}
\affiliation{European XFEL, Holzkoppel 4, 22869 Schenefeld, Germany}
\author{Claudio Masciovecchio}
\affiliation{Elettra-Sincrotrone Trieste, SS 14 km 163,5 in AREA Science Park, 34149, Trieste, Italy.}
\author{Riccardo Mincigrucci}
\affiliation{Elettra-Sincrotrone Trieste, SS 14 km 163,5 in AREA Science Park, 34149, Trieste, Italy.}
\author{Ettore Paltanin}
\affiliation{Elettra-Sincrotrone Trieste, SS 14 km 163,5 in AREA Science Park, 34149, Trieste, Italy.}
\author{Filippo Bencivenga}
\affiliation{Elettra-Sincrotrone Trieste, SS 14 km 163,5 in AREA Science Park, 34149, Trieste, Italy.}


\begin{abstract}
\noindent
We crossed two femtosecond extreme ultraviolet (EUV) pulses in a $\beta -Ga_2O_3$ (001) single crystal to create transient gratings (TG) of light intensity with sub-100 nm spatial periodicity. The EUV TG excitation launches phonon modes, whose dynamics were revealed via the backward diffraction of a third, time-delayed, EUV probe pulse. In addition to the modes typically observed in this kind of experiment, the phase-matching condition imposed by the TG, combined with the sharp penetration depth of the EUV excitation pulses, permitted to generate and detect phonons with a wavevector tangibly larger ($\approx$ 1 nm$^{-1}$) than the EUV TG one, via stimulated Brillouin back-scattering (SBBS) of the EUV probe. While SBBS of an optical probe was reported in previous EUV TG experiments, the extension of SBBS to short wavelength radiation can be used as a contact-less experimental tool for filling the gap between the wavevector range accessible through inelastic hard X-ray and thermal neutron scattering techniques, and the one accessible through Brillouin scattering of visible and UV light.
\end{abstract}

\maketitle



\begin{figure*}
\centering
\centering\includegraphics[width=\linewidth]{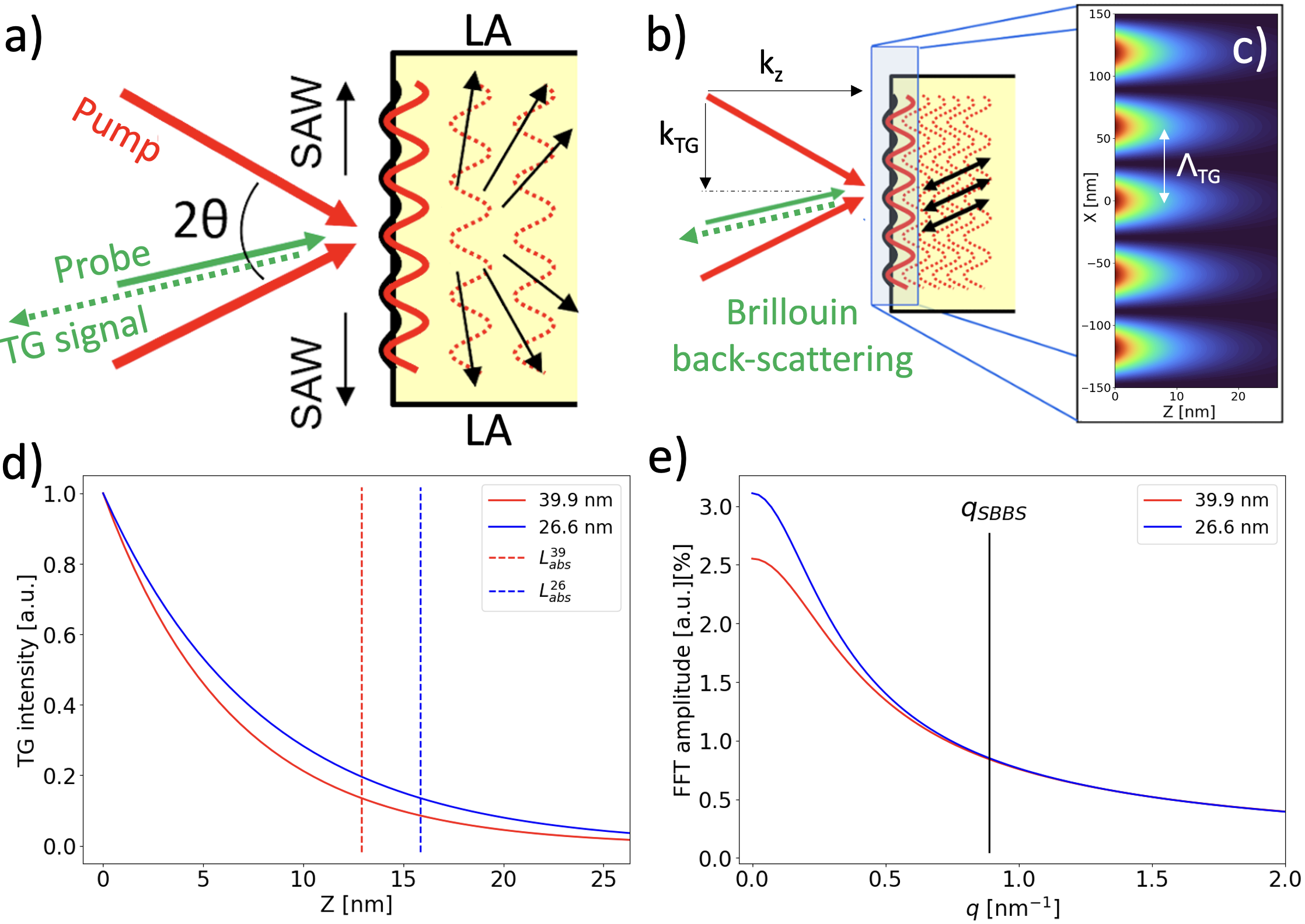}
\caption{The schematic configuration of the EUV TG setup is shown in a). Two pump beams, having the same wavelength, $\lambda$, are spatially and temporally overlapped on the sample with a crossing angle equal to $2\theta$, generating an intensity grating with modulation along $q_{\text{TG}}$ (parallel to the surface). Such a spatially periodic excitation with period $\Lambda_{\text{TG}}$ launches surface acoustic waves (SAW) along $q_{\text{TG}}$ and surface skimming longitudinal acoustic (LA) modes in the sub-surface region. These modes are probed by a third, time-delayed pulse and detected in backward diffraction geometry. Panel b) illustrates how the setup enables the simultaneous detection of the stimulated Brillouin back-scattered signal, which is co-axial with the TG signal but generated by the intensity variation along $q_z$. The excitation intensity of the EUV TG (for the 26.6/13.3 nm pump/probe pair) is depicted in panel c). Panel d) displays a longitudinal slice of panels c) and its counterpart at 39.9 nm. This steep modulation launches acoustic waves in a broad range of $q$, which is represented by the FT along z of the excitation intensity profiles (see panel e)).}
\label{fig:fig_1}
\end{figure*}

\noindent
Studying thermal and vibrational dynamics in nanoscale materials is critical for advancing the technological applications of faster, more efficient and more compact nanoelectronic devices, such as smartphone and computer chips, as well as for thermal barrier
coatings \cite{qian2021phonon}, heat-assisted magnetic recording \cite{rausch2014heat}, nano-enhanced photovoltaics and thermoelectric energy conversion, to name a few. To achieve this, layer upon layer of very thin films are often used, with impurities added to tailor their function \cite{frazer2020full}. However, the complex structure of these materials makes it challenging to predict and characterise their thermoelastic properties. 
Material properties such as elasticity, thermal conductivity and heat capacity are mostly determined by collective lattice dynamics that exhibit strong length-scale dependencies, which can drastically differ when the spatial dimensions reduce from macroscopic to microscopic scales, \textit{i.e.}, to sizes comparable with the characteristic length scales of nanostructures.\\
\hspace*{3mm} Over the years, an obstacle to the full description of thermoelastic responses in the 10s of nm length-scale was given by the lack of experimental techniques capable of accessing such range \cite{bencivenga2009fel} without the requirement of
modifying or physically touching the sample. This inherently introduces limitations in the experiment design and complicates data interpretation. Collective lattice dynamics in condensed matter at wavevector $q > 1$ nm$^{-1}$ can be measured by inelastic scattering of hard X-ray and thermal neutron, while Brillouin scattering and optical transient grating (TG) can be used for $q < 0.1$ nm$^{-1}$. The intermediate $q = 0.1-1$ nm$^{-1}$ is hardly accessible, despite efforts to expand the capabilities of Brillouin spectroscopy in the UV range \cite{bencivenga2012high} and for improving the performance of X-ray spectrometers \cite{shvyd2014high}. In addition, these spectroscopic methods are inherently limited by the instrumental resolution when measuring narrow lines, \textit{i.e.} long dynamics. This limitation does not affect time-domain techniques, such as picosecond ultrasonics and time-domain thermoreflectance. In these techniques, metal films or other nanostructures are fabricated on the sample for transducing an ultrafast optical excitation in a short wavelength thermoelastic perturbation \cite{legrand2016direct,chou2019long}. However, this intrinsically modifies the sample under investigation.\\
\hspace*{3mm} The advent of free-electron laser (FEL) sources has recently permitted the usage of extreme ultraviolet (EUV) pulses for extending the TG approach to shorter wavelengths, \textit{i.e.} in the $10-100$ nm range, enabling the excitation and probing of nanoscale thermoelasticity in a contact-less fashion \cite{bencivenga2019nanoscale,foglia2023extreme}. The EUV TG approach has been pioneered at the FERMI FEL (Trieste, Italy) with the dedicated endstation TIMER \cite{mincigrucci2018advances,bencivenga2023extreme}, capable of incisively and selectively studying bulk and surface phonons \cite{maznev2021generation}, thermal transport kinetics \cite{naumenko2019thermoelasticity,bencivenga2019nanoscale} and magnetic dynamics \cite{ksenzov2021nanoscale}.\\
\hspace*{3mm} In this paper, we exploit EUV TG to probe acoustic phonons in $\beta-Ga_2O_3$. In particular, by taking advantage of the phase-matching conditions imposed by the nanoscale EUV TG, we demonstrated the possibility to detect stimulated Brillouin back-scattering (SBBS) from an EUV pulse at 13.3 nm wavelength. This enabled us to probe the dynamics of phonon modes with a wavelength as short as $\approx$ 6 nm. The employed sample was an Mg-doped $\beta-Ga_2O_3$ (001)-oriented bulk crystal with monoclinic structure (space group C2/m), obtained from the Czochralski method at the Leibniz-Institut für Kristallzüchtung \cite{galazka2016scaling}. The excellent surface quality and well-known elastic parameters made this sample adapted for the present EUV TG experiment. 



\begin{figure*}
\centering
\centering\includegraphics[width=\linewidth]{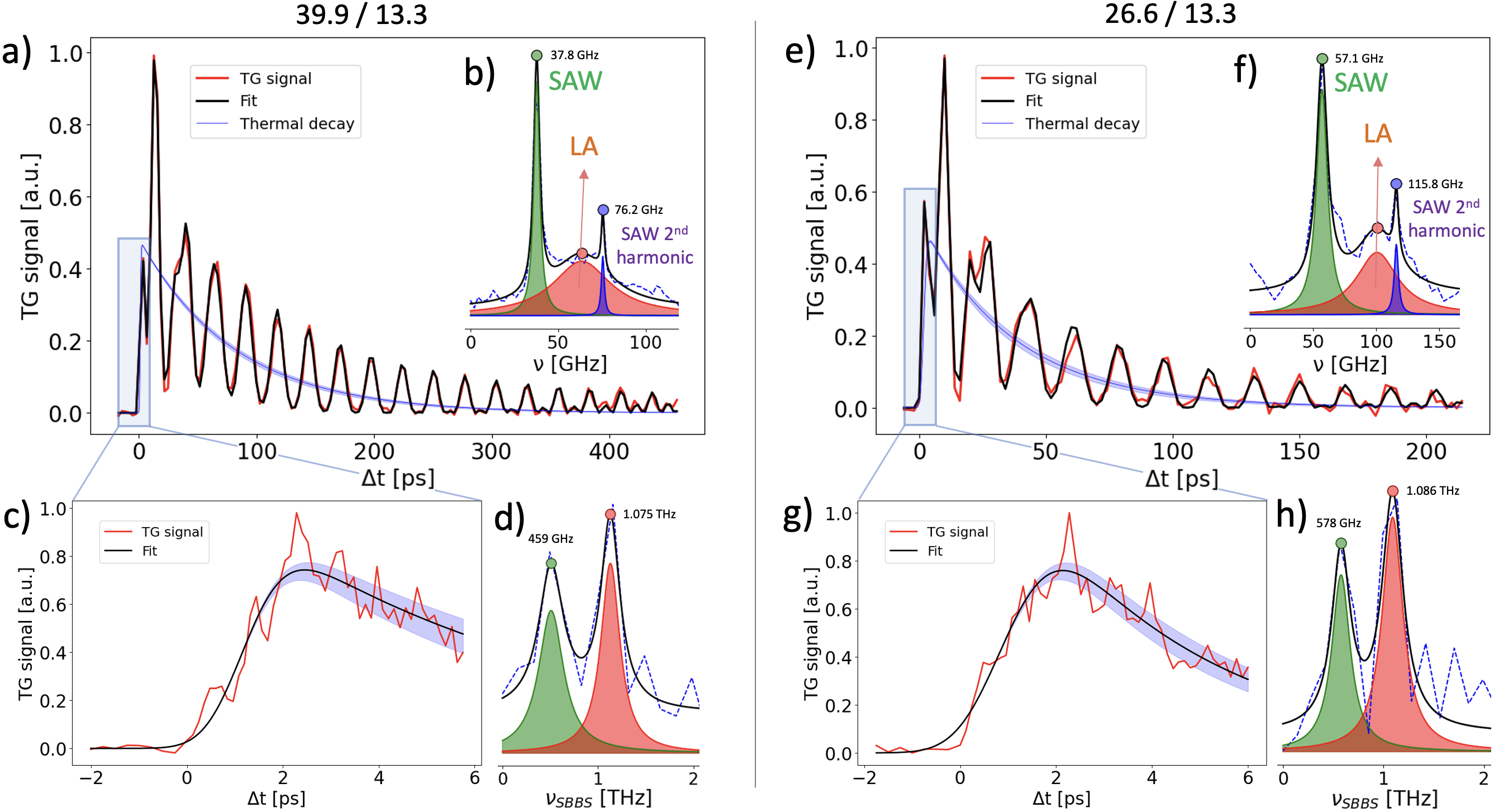}
\caption{Panel a) and e) show the long timescale dynamic of the EUV TG signal. The slow decay is displayed in blue; the associated error is represented by the thickness of this line. The total fit of the signal, guided by the FT in b) and f), is shown in black. Panel c) and g) present a zoom-in of the first few ps, showing SBBS oscillations. The FT, obtained after removing the slow signal variation (black line in c) and g)), is reported in d) and h), showing two modes.}
\label{fig:fig_2}
\end{figure*}

TG is a third-order non-linear optical technique (four-wave-mixing), wherein two pulses of equal wavelength $\lambda$ (referred to as pumps) are temporally and spatially
overlapped on the sample at a crossing angle of $2\theta$. The interference between these two pulses, assuming parallel polarization of the beams, induces a spatial modulation in the intensity of light. This modulation exhibits a periodicity $\Lambda_{\text{TG}}=\lambda/(2\sin \theta)$; see Figs. \ref{fig:fig_1}a)-\ref{fig:fig_1}b). Such a patterned excitation acts as a transient diffraction grating for a third variably-delayed pulse (probe), with wavelength $\lambda_{\text{pr}}$, giving rise to a fourth pulse: the diffracted beam (signal).\\
\hspace*{3mm} The experiment was performed at the TIMER beamline at the FERMI FEL, which is described in detail elsewhere \cite{mincigrucci2018advances}. Two time-coincident $\approx$ 60 fs (FWHM) EUV pulses were crossed on a crystalline $\beta-Ga_2O_3$ (001) sample at the angle $2\theta=27.6^{\circ}$ (set with 2$\%$ accuracy), generating a transient grating in the [100] direction. Two values of $\lambda$ were used: 39.9 nm and 26.6 nm, resulting in corresponding grating periods of $\Lambda_{\text{TG}}\approx$ 84 nm and $\approx$ 56 nm, respectively. In the following, we will refer to the 39.9 nm and 26.6 nm pump-related quantities with the superscript $^{39}$ and $^{26}$, respectively. The probe pulse ($\approx$ 40 fs FWHM) impinged on the sample with an angle of 4.6$^{\circ}$, and $\lambda_{\text{pr}}=13.3$ nm (hereafter denoted as $^{13}$). The backwards-diffracted signal beam was collected by a EUV mirror and detected by a CCD camera, as outlined in \cite{maznev2021generation}. The beamline is designed to satisfy the TG phase matching conditions at the Bragg angle (\textit{i.e.} $\theta_i = \theta_o = \sin^{-1}(\lambda_{\text{pr}}/2\Lambda_{\text{TG}})$; being $\theta_i$ and $\theta_o$ the incidence and diffraction angles of the probe beam, respectively) for $\lambda=3\lambda_{\text{pr}}$. However, since the excitation light is absorbed in a subsurface layer shorter than $\Lambda_{\text{TG}}$ (the absorption lengths of the pumps are: $L_{\text{abs}}^{39} \sim 12.9$ nm and $L_{\text{abs}}^{26} \sim 15.9$ nm), phase matching conditions are relaxed. In this case only the wavevector component parallel to the sample surface ($q_{\text{TG}}^{39}=2\pi/\Lambda^{39}_{\text{TG}}\approx 0.075$ nm$^{-1}$, and $q_{\text{TG}}^{26}\approx 0.113$ nm$^{-1}$) is well-defined \cite{foglia2023extreme}, while the component perpendicular to the surface ($q_z$) results in a broad spectrum; see Fig. \ref{fig:fig_1}. Therefore, acoustic waves with a well-defined wavevector equal to $q_{\text{TG}}$ (parallel to the surface) are launched. In contrast, waves with a broad spectrum in $q_z$ are generated along the z-direction. However, as shown in Ref \cite{maznev2018generation}, only two values of $q_z$ satisfy the TG phase-matching conditions \textit{i.e.}, $q_z=0$, which yields a signal in the forward direction, and

\begin{equation}
\label{eq:eq_brill_2}
q_z=2k\sqrt{1-q_{\text{TG}}^2/4k^2}
\end{equation}

\noindent
yielding a back-scattered signal, that encodes the dynamics of SBBS modes. Here, $k=2\pi n /\lambda_{\text{pr}}$ is the wavevector of the probe in the medium, where $n$ is the refractive index at $\lambda_{\text{pr}}$. Thus, the modulus of the acoustic wavevector for the SBBS signal is:

\begin{equation}
\label{eq:eq_brill}
q_{\text{SBBS}}=\sqrt{q_z^2+q_{\text{TG}}^2}=2k=\frac{4\pi n}{\lambda_{\text{pr}}} 
\end{equation}

\noindent
which is independent of $q_{\text{TG}}$ and is collinear with the backward diffracted signal from the TG.\\
\hspace*{3mm} Therefore, under the current experimental conditions, the combination of the sharp penetration depth of the EUV TG pump in the material and the short wavelength EUV probe, enables the excitation and detection of phonons with $q$ as large as $q_{\text{SBBS}}\approx$ 1 nm$^{-1}$ (Fig. \ref{fig:fig_1}b). To further illustrate the excitation mechanism, Fig. \ref{fig:fig_1}c displays the EUV TG generated on the sample in the 26.6/13.3 configuration plotted against the ($x$,$z$) coordinates, taking into account the finite value of $L^{26}_{\text{abs}}$. We note that the modulation along $x$  extends in a much larger range, comparable with the width ($\text{FWHM}_{x}\approx 100$s of $\mu$m) of the excitation pulses. This is what we usually call TG. Additionally, there is a steep gradient along $z$. Such gradient launches acoustic waves in a broad range of $\Delta q$ (roughly extending up to $\sim 2\pi /L^{\text{pump}}_{\text{abs}}$). This is represented in Fig. \ref{fig:fig_1}e as the Fourier transform (FT) along $z$ of the EUV TG intensity profile shown in Fig. \ref{fig:fig_1}d. In an excitation scheme relying on a single EUV pump, there is no capability to selectively choose a specific phonon wavevector along the $z$-axis. However, in the current scenario, the phase matching condition imposed by the EUV TG (see Eq.\ref{eq:eq_brill}) selects a specific wavevector phonon with $q_{\text{SBBS}}\sim$ 1 nm$^{-1}$ from the wide range of available phonons This is illustrated by the vertical segment in Fig. \ref{fig:fig_1}e.\\
\hspace*{3mm} We detected the EUV TG signal by varying the time delay ($\Delta t$) between the EUV TG excitation and the probe pulse. Measurements were conducted at both long timescales (Figs. \ref{fig:fig_2}a and \ref{fig:fig_2}e) and short timescales (Figs. \ref{fig:fig_2}c and \ref{fig:fig_2}g). As expected, at long timescales the overall signal is characterized by a slow decay, which can be attributed to the thermal relaxation of the EUV TG, modulated by phonon oscillations \cite{bencivenga2019nanoscale,maznev2021generation}. After a few oscillations, these modulations become highly regular. For larger $\Delta t$ values, when the slow relaxation is decayed, double-frequency oscillations become visible, indicating the long-living nature of this dominant mode \cite{maznev2021generation}. Conversely, the irregular shape of the initial oscillations suggests the presence of additional dynamics that damps out after some 10s of ps. The EUV TG data obtained at short timescales (Figs. \ref{fig:fig_2}c and \ref{fig:fig_2}g) were sampled with finer steps and exhibit modulations at significantly higher frequencies. These higher-frequency modulations are compatible with the previously mentioned mixing between the SBBS signal and the backward diffracted signal from the EUV TG.\\
\hspace*{3mm} In order to quantitatively describe the waveforms at both long (blue line in Fig. \ref{fig:fig_2}a and \ref{fig:fig_2}e) and short timescale (black line in Fig. \ref{fig:fig_2}c and \ref{fig:fig_2}g) an initial fitting procedure was conducted using Eq. \ref{eq:eq_1}:

\begin{equation}
I(t) = \Bigl|\frac{1}{2}\Bigl[1+\textit{erf}\Bigl(\frac{\Delta t}{\sigma}\Bigr)\Bigr] \cdot A\, e^{-\frac{\Delta t}{\tau}}\Bigr|^2,
\label{eq:eq_1}
\end{equation}

\noindent
where the \textit{erf} function accounts for a sudden rise of the signal (with $\sigma$ representing the width of the rise), followed by an exponential decay with a time constant $\tau$. Subsequently, FTs were computed on the differences between the measured traces and their respective exponential fits. The obtained results are illustrated in Figs. \ref{fig:fig_2}b, \ref{fig:fig_2}d, \ref{fig:fig_2}f and \ref{fig:fig_2}h.\\
\hspace*{3mm} The FTs of the long timescale waveforms present a well-defined mode and its second harmonic, plus a weaker and spectrally broader feature. All frequencies in these FTs vary proportionally to $q_{\text{TG}}$, as depicted in Figs. \ref{fig:fig_2}b) and \ref{fig:fig_2}f). The presence of this broad feature confirms the existence of a damped mode, which predominantly affects the initial portion of the waveform, as already evident from the raw data.\\
\hspace*{3mm} To comprehensively describe the signal, the complete fitting procedure incorporated these two vibrational modes, specifically a damped sinusoidal term and an undamped sinusoidal term:

\begin{equation}
\label{eq:eq_2}
\begin{split}
I(t) =& \Bigl|\frac{1}{2}\Bigl[1+\text{erf}\Bigl(\frac{\Delta t}{\sigma}\Bigr)\Bigr] \cdot \Bigl[A\, e^{-\frac{\Delta t}{\tau}} + \\ & - A_{\text{SAW}}\sin(2\pi\,\nu_{\text{SAW}}\,\Delta t + \phi_{\text{SAW}}) + \\ & - A_{\text{LA}}\sin(2\pi\,\nu_{\text{LA}}\,\Delta t + \phi_{\text{LA}})\, e^{-\frac{\Delta t}{\tau_{\text{LA}}}}\Bigr]\Bigr|^2
\end{split}
\end{equation}

\begin{figure}[!ht]
\centering
\centering\includegraphics[width=8.7cm]{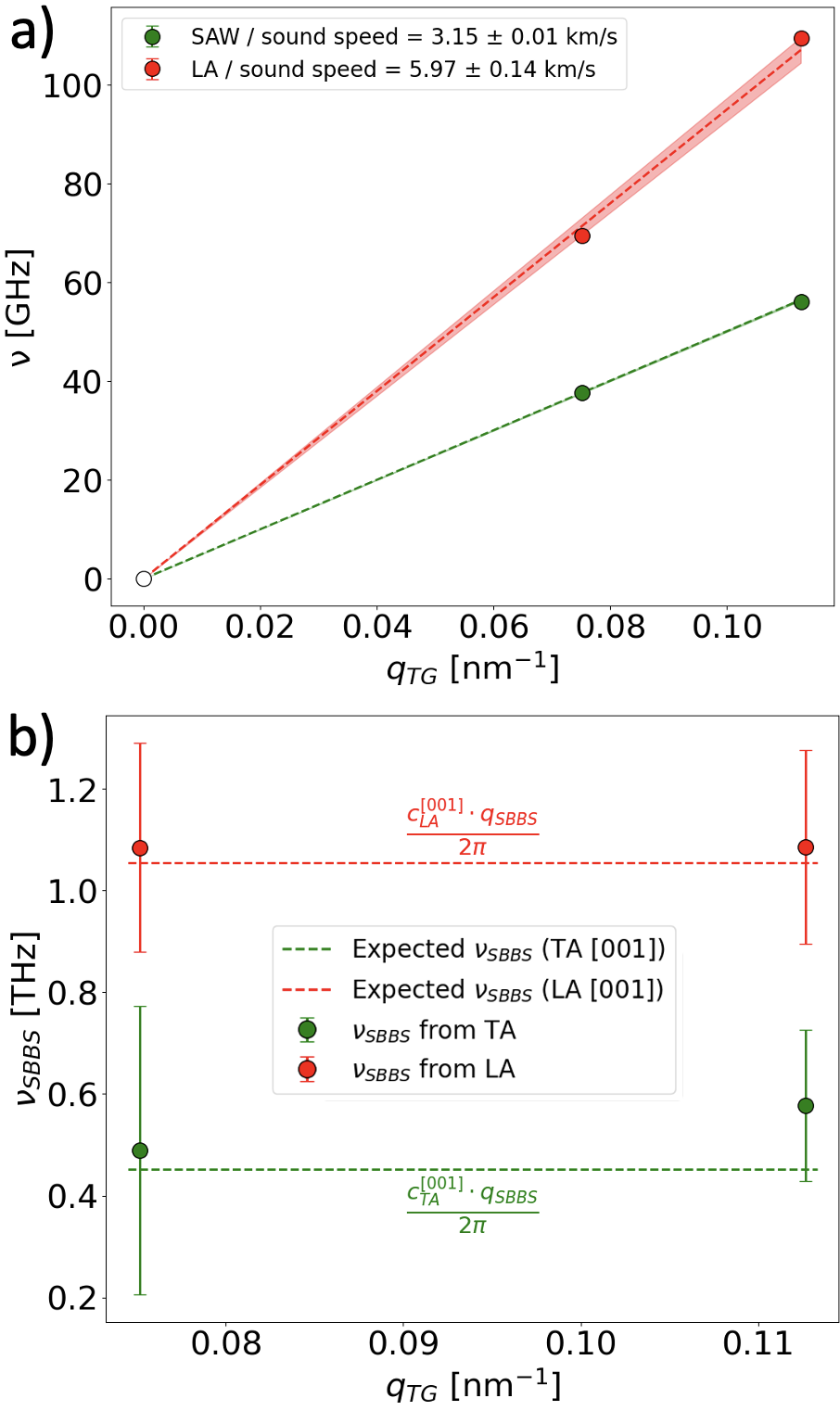}
\caption{The modes visible in the long timescale signal are presented in panel a). Both SAW and LA show a linear dependence on $q_{\text{TG}}$. Panel b) displays the SBBS frequencies which do not vary with $q_{\text{TG}}$, since in the present case the dependence of $q_{\text{SBBS}}$ on $q_{\text{TG}}$ can be neglected. Dashed lines are the expected frequencies given by $ \nu_{\text{SBBS}}=c_{\text{LA/TA}}\cdot q_{\text{SBBS}}/2\pi$. The shift of the TA $\nu_{\text{SBBS}}$ at $q_{\text{TG}}^{26}$ can be due to the $q_{\text{SBBS}}$ direction pointing further away from the [001] direction which shows a higher frequency shift with respect to the LA mode \cite{mengle2019vibrational}}
\label{fig:fig_3}
\end{figure}

\noindent
The resulting best-fit results are reported as black lines in Figs. \ref{fig:fig_2}a and \ref{fig:fig_2}e. All parameters and errors mentioned further below have been obtained using Eq. \ref{eq:eq_2}. The values obtained from the preliminary fitting of the EUV TG signal with Eq. \ref{eq:eq_1} and from the FTs were used as an initial guess for fitting the data with Eq. \ref{eq:eq_2}.\\
\hspace*{3mm} The results concerning the oscillation frequencies are shown in Fig. \ref{fig:fig_3}a. The undamped mode is compatible with a Surface Acoustic Wave (SAW), which exhibits a linear dispersion relation as a function of $q_{\text{TG}}$. From the slope of such liner dispersion a value for the sound velocity of $c_{\text{SAW}}^{[100]} = 3.15 \pm 0.01$ km/s is obtained. This value is close to the estimated velocity of 3.24 km/s, as evaluated by using the transverse acoustic (TA) phonon velocity $c_{\text{TA}}^{[100]}=3.57$ km/s \cite{mengle2019vibrational} and the Poisson's ratio $\nu_p=0.2$ \cite{zheng2021young} of $\beta-Ga_2O_3$ [100], through the relation $c_{\text{SAW}}\approx c_{\text{TA}}\cdot (0.862+0.14\nu_p)/(1+\nu_p)$ \cite{freund1998dynamic}. SAW modes represent long-lived coherent surface displacements characterized by mechanical energy confined to the surface. In the employed backward diffraction geometry, these modes are expected to be the dominant contribution to the EUV TG signal, as observed experimentally.\\ 
\hspace*{3mm} The damped mode also presents a liner dispersion with a velocity $c_{\text{LA}}^{[100]}=5.97 \pm 0.14$, which is similar to the expected value (6.18 km/s) for longitudinal acoustic (LA) phonons \cite{mengle2019vibrational}. Such marginal deviations between the expected and observed velocities may arise from factors such as slight misalignment of the sample relative to the [100] crystallographic direction, sample heating caused by the FEL, or the $10^{^\circ}$ tilt in the $(x,y)$-plane, necessary for collecting the backward diffracted signal \cite{mincigrucci2018advances}.
\\
\hspace*{3mm} Surface-skimming LA modes and, more in general, bulk waves are expected in these types of TG experiments \cite{zoubkova2021transient,januvsonis2016transient}, although they do not contribute significantly in the employed geometry and are often disregarded. Furthermore, at these $q$ values bulk modes are not expected to show tangible damping in the probed $\Delta t$ range. However, EUV TG data indicate a quite fast decay time, \textit{i.e.}: $\tau^{39}_{\text{LA}} = 22.5 \pm 1.5$ ps and $\tau^{26}_{\text{LA}} = 17.6 \pm 1.5$ ps, which is compatible with the broad feature observed in the FT (see Fig. \ref{fig:fig_2}b and \ref{fig:fig_2}f). The finite decay time can be explained by the fact that we are observing a thin region below the surface, with thickness $\approx L_{\text{abs}}^{13} \sim 26.3$ nm $ < \Lambda_{\text{TG}}$, and the excitation intensity steeply varies along the sample depth. Consequently, LA modes are strongly influenced by the surface and manifest as leaky waves, such as surface-skimming longitudinal waves, which rapidly transfer mechanical energy away from the subsurface region toward the bulk.\\ 
\hspace*{3mm} The FTs of the short timescale waveforms exhibit two peaks (Figs. \ref{fig:fig_2}d and \ref{fig:fig_2}h) located at considerably higher frequencies compared to SAW and LA modes. Furthermore, these peaks do not show dispersion vs $q_{\text{TG}}$, as shown in Fig. \ref{fig:fig_3}b. This behaviour is indeed expected from the SBBS of the EUV probe, since the phonon wavevector is given by $q_{\text{SBBS}}$ and in this specific case the dependence on $q_{\text{TG}}$ can be neglected (see Eq. \ref{eq:eq_brill}). The absence of dispersion vs $q_{\text{TG}}$ of phonon modes detected via SBBS does not imply that they do not exhibit dispersion; rather, it indicates that the changes in $q_{\text{TG}}$ allowed under the specific experimental conditions were not sufficient to significantly alter $q_{\text{SBBS}}$. A more effective approach to modifying $q_{\text{SBBS}}$ would be to vary $\lambda_{\text{pr}}$, as in this case, $q_{\text{SBBS}} \propto \lambda_{\text{pr}}^{-1}$ (see Eq. \ref{eq:eq_brill}).\\
\hspace*{3mm} On the other hand, the observed frequencies ($\nu_{\text{SBBS}}$, as extracted from the FT) match with the ones expected by considering the sound velocities of TA ($c_{\text{SAW}}^{[001]}=4.01$ km/s) and LA ($c_{\text{LA}}^{[001]}=7.55$ km/s) modes along the relevant crystallographic direction \cite{mengle2019vibrational}; see Fig. \ref{fig:fig_3}b. Indeed, the LA mode detected via SBBS propagates along $q_{\text{SBBS}}$, which means with a small tilt angle ($\phi^{39}=4.8^{\circ}$ and $\phi^{26}=7.2^{\circ}$) with respect to $q_z$, \textit{i.e.,} essentially towards the bulk of the sample ([001]). This is a different crystallographic direction with respect to the leaky LA mode detected at long timescales (see Figs. \ref{fig:fig_2}a and \ref{fig:fig_2}e) which essentially propagates beneath the surface ([100]) with wavevector $q_{\text{TG}}\ll q_{\text{SBBS}}$. However, since the employed setup did not allow precisely selecting crystallographic directions, such modes have to be regarded as quasi-LA and quasi-TA. It is worth mentioning that Brilluoin back-scattering from quasi-TA modes can be observed in monoclinic crystals, exhibiting signal amplitudes (in the optical regime) comparable to those from quasi-LA modes \cite{kasprowicz2007elastic,vacher1972brillouin,Note}. However, while EUV Brillouin scattering reasonably relies on the same selection rules as in the optical regime, the signal amplitude may differ due to potential wavelength-dependent variations in the photoelastic constants. Most likely, the modes associated with larger density variations provide stronger signals, as the EUV refractive index (far from core-hole resonances) primarily depends on density \cite{bencivenga2023extreme}. Nevertheless, further experiments beyond the scope of this study are required to investigate these aspects.\\
\hspace*{3mm} The combination of the sharp penetration depth of EUV excitation pulses and the phase-matching conditions imposed by the EUV TG permitted the detection of stimulated backscattered Brillouin oscillations with a wavevector as large as $\approx 1$ nm$^{-1}$. This wavevector range overlaps with the lower limit of the wavevector range covered by inelastic scattering of hard X-ray and thermal neutrons. In this case, the limitations on the $q_{\text{SBBS}}$ and the SBBS signal come from the wavelength of the probe, rather than from the EUV TG periodicity (see Eq. \ref{eq:eq_brill}). This limit can be straightforwardly overcome by using a shorter probe wavelength, that can be envisioned extending all the way to the X-ray spectral range \cite{rouxel2021hard}. This would provide a longer penetration depth and an increased range in $q_{\text{SBBS}}$.\\
\hspace*{3mm} Furthermore, the described approach also allowed for the detection of high-frequency surface acoustic waves and longitudinal acoustic phonons propagating below the surface, without the need for nanofabrication and in a broad range of materials. In fact, unlike optical laser excitation, EUV photons are highly absorbed by any materials. The current setup at FERMI already makes it possible to conduct transient grating measurements at grating periods as short as 24 nm \cite{bencivenga2015four,foglia2023extreme}, and a further extension down to approximately 10 nm is feasible, pushing the SAW frequency close to the THz region and $q_{\text{SBBS}}$ above 1 nm$^{-1}$.
\begin{acknowledgments}
\noindent
The authors thank Z. Galazka from Leibniz-Institut für Kristallzüchtung for providing the $\beta-Ga_2O_3$ (001) sample and Alexei Maznev (MIT, Boston) for useful discussions. E. P. acknowledges funding from the European Union’s Horizon 2020 research and innovation programme under the Marie Skłodowska-Curie grant agreement No 860553.
\end{acknowledgments}


\bibliography{apssamp}

\end{document}